\documentclass[sigconf, nonacm]{acmart}
\usepackage{pvldb}
\usepackage{makecell}
\usepackage{multirow}
\usepackage{enumitem}
\usepackage[table,dvipsnames]{xcolor}
\usepackage{chngcntr}
\newcommand{\rev}[1]{{\color{black}#1}}

\makeatletter
\let\c@table\c@figure

\makeatother
\renewcommand\vldbdoi{XX.XX/XXX.XX}
\renewcommand\vldbpages{XXX-XXX}
\renewcommand\vldbavailabilityurl{} 

\begin{document}
\title[Recommendation Systems for Exploratory Data Tasks]
	{Recommendation Systems for Exploratory Data Tasks [Vision]}

\author{Anna Fariha}
\affiliation{%
  \institution{University of Utah}
  \city{Salt Lake City}
  \state{Utah}
  \country{USA}
}
\email{afariha@cs.utah.edu}

\author{Whanhee Cho}
\affiliation{%
  \institution{University of Utah}
  \city{Salt Lake City}
  \state{Utah}
  \country{USA}
}
\email{whanhee.cho@utah.edu}

\begin{abstract} 

\looseness-1 \underline{E}xploratory \underline{D}ata \underline{T}asks (EDTs)
involve iteratively steering workflows as users refine their subjective goals
based on newly emerging insights. EDTs are performed by millions of users with
varying levels of expertise to understand unfamiliar data, discover trends, and
identify evidence that informs critical decision-making. However, a key
challenge in EDTs is the enormous space of possible \emph{actions} that one can
take at each step: users struggle to choose among thousands of joins,
transformations, and aggregations, leading to ``exploration paralysis''.
Because EDT workflows are interconnected, each choice shapes subsequent
exploration, and suboptimal choices can lead to inefficiency, missed insights,
confirmation bias, and incomplete coverage. This calls for intelligent
recommendations that efficiently guide users toward optimal EDT actions.

\begin{sloppypar} We envision \emph{recommendation as a core capability of data
systems}, proactively guiding users toward promising actions that lower the
barrier to EDTs. \rev{This capability becomes even more critical as AI agents
emerge as a new class of users that perform EDTs on behalf of humans.} EDT
recommendation is challenging because the action space is combinatorial and
actions are data-dependent, which require costly materialization. Moreover,
recommendation often involves bundles or sequences of actions across
interdependent tasks, requiring coordination across tasks. In this paper, we
present our vision of EDT recommendation systems along two axes: single-task
vs.\ multi-task settings and single-action vs.\ multi-action recommendations.
We outline a research agenda that progresses from recommending individual EDT
actions to constrained bundles and sequences of actions, and ultimately to
coordinated recommendations across interconnected EDTs. We identify research
directions for incorporating various contexts (user, data, task, and
ecosystem), addressing efficiency challenges, and coordinating across tasks.
\end{sloppypar}

\end{abstract}

\maketitle

\vldbtopmatter

\section{Introduction}

\noindent In today's world, users are overwhelmed by abundant content (e.g.,
products, movies, news), making recommendation essential for information
filtering. Recommendation systems assist users in discovering relevant items
from a large pool of candidates---by analyzing past behavior, preferences, and
contextual signals---while striking a balance between exploration (discovery)
and exploitation (personalization). Beyond filtering content, recommendation
can also guide users through complex tasks by suggesting next actions. Such
task recommendations support rapid onboarding for novices and improve
efficiency for experts. While recommendation systems are well-studied in
content domains, their use in data management tasks remains underdeveloped.

\looseness-1 We focus on \emph{Exploratory Data Tasks
(EDTs)}~\cite{10.14778/3773731.3773746}, where users iteratively steer
workflows, refining subjective goals as new insights emerge. For instance, data
analysts choose which data subsets to drill into based on emerging patterns,
and database administrators reason about diagnostic queries and logs to guide
subsequent debugging steps. Similar examples arise across numerous applications
such as exploratory data analysis~\cite{DBLP:conf/icde/JoglekarGP16},
summarization~\cite{cho2025data}, insight
discovery~\cite{DBLP:journals/pvldb/XingWJ24},
explanation~\cite{DBLP:journals/pvldb/ShragaM23}, and
visualization~\cite{DBLP:journals/pvldb/VartakRMPP15}. As data becomes central
to decision-making, EDTs are ubiquitous across users of all expertise levels.
\rev{Moreover, the users of data systems now include AI agents that perform
EDTs on behalf of humans, taking exploratory actions at machine speed.}
However, users must repeatedly choose the next action from thousands of options
such as joins, aggregations, transformations, etc. Suboptimal choices lead to
(1)~\emph{inefficiency}, wasting human time, effort, and computational
resources on futile directions; (2)~\emph{confirmation bias}, arising from an
unprincipled exploration process; and (3)~\emph{missed insights}, due to
portions of the data remaining unexplored, which may lead to incorrect
conclusions and adversely affect critical decision-making. \rev{These issues
are amplified for agents: without guidance, agents explore by generating and
executing actions through trial and error, which significantly increases the
load on the data systems.}

Exploration is an iterative process and choosing the next best action at each
step is challenging due to two key reasons. First, unlike fully observable
entities such as tuples in a database, EDT actions are merely \emph{programs
parameterized by the underlying data}. Thus, the utility of an action is
difficult to assess without materializing it over the data. Second, the space
of possible EDT actions is often \emph{combinatorial} (Example~\ref{ex:one}),
which overwhelms users with too many choices, leading to ``exploration
paralysis''. Programmatically enumerating and materializing this vast space of
candidate actions is computationally prohibitive, motivating the need for
\emph{recommendations} toward choosing optimal EDT actions.

\begin{example}[Combinatorial search space of EDT actions]\label{ex:one}
	 \looseness-1 Consider a business dataset with the exploratory goal of
	 identifying discrepancies across groups under some (unknown) data
	 partitioning, by computing various \textcolor{ForestGreen}{aggregated}
	 \textcolor{Cyan}{attributes} such as
	 \textcolor{ForestGreen}{$\mathtt{AVG}$}\,(\textcolor{Cyan}{$\mathtt{salary}$})
	 or \textcolor{ForestGreen}{$\mathtt{SUM}$}\,(\textcolor{Cyan}{$\mathtt{cost}$})
	 grouped by various \textcolor{OrangeRed}{attribute combinations} such as
	 \textcolor{OrangeRed}{$\mathtt{gender}$ \& $\mathtt{education}$} or
	 \textcolor{OrangeRed}{$\mathtt{region}$ \& $\mathtt{year}$}. Even with only
	 \textcolor{ForestGreen}{$5$ aggregate functions} over \textcolor{Cyan}{$20$
	 numeric attributes} and \textcolor{OrangeRed}{$20$ grouping attributes}, and
	 restricting queries (actions) to only 2 aggregate and 2 grouping attributes,
	 the number of possible choices is $5^2 \cdot \binom{20}{2} \cdot
	 \binom{20}{2} \approx$ 1 million!
\end{example}

\looseness-1 User perception of exploratory findings is highly subjective and
must inform action utility, making \emph{personalization} essential during EDT
recommendation (Example~\ref{ex:two}). Yet, existing data systems provide
limited support for EDT recommendations, leading to adoption of AI chatbots:
users are exposed to unvalidated suggestions driven by surface-level patterns
or popularity signals, not principled notions of exploration such as data
coverage and bias avoidance. While incorrect recommendations in entertainment
are inconsequential, they can have catastrophic consequences in high-stakes
decision-making. This critical risk around AI safety further motivates the need
for principled foundations for EDT recommendation.

\begin{example}[Need for personalization]\label{ex:two}
    \looseness-1 Consider two users exploring the dataset of
    Example~\ref{ex:one} with the same intent of finding discrepancies. One is
    an investigative journalist and finds evidence of discrimination across
    demographic groups (e.g., gender-based $\mathtt{salary}$ disparities)
    insightful. The other is a company manager and finds discrepancies in
    $\mathtt{resource}$ $\mathtt{utilization}$ across $\mathtt{team}$,
    $\mathtt{region}$, or $\mathtt{year}$ insightful. The intent and data
    remain the same, yet the perceived utility of an action differs
    significantly between users.
\end{example} 

While identifying a single action is already challenging, the problem is
exacerbated when users must identify a coherent \emph{set} or \emph{sequence}
of actions in a \emph{multi-task} setting, with dependencies across tasks, such
as data discovery, followed by wrangling, and concluding with summarization.
The complexity of this setting requires new theoretical and algorithmic
foundations for unified recommendation across multiple, interdependent EDTs
(Example~\ref{ex:three}).

\begin{example} [Unified recommendation in a multi-task setting]\label{ex:three}
   During data discovery, a recommender suggests a temperature table from a
   data lake. During wrangling, a second recommender suggests forward-fill
   imputation for missing temperature readings based on the time-series
   semantics inferred during discovery. At the summarization step, a third
   recommender avoids trends over temperature, recognizing its reduced
   reliability due to imputation.
\end{example}

\smallskip\noindent\textbf{Our vision.} We envision EDT recommendation as a
core capability of data systems, satisfying three key desiderata:

\begin{enumerate}[leftmargin=*, labelsep=1pt]

    \item \emph{high utility}, by ensuring relevant, semantically valid,
    personalized, context-aware, and adaptive recommendations;
	
    \item \emph{practical efficiency}, by managing the prohibitive
    computational cost of enumerating a combinatorial number of candidate
    actions and their materialization; and

   \item \emph{comprehensive} support across single-task single-action,
   single-task multi-action, and multi-task multi-action settings.

\end{enumerate}


\smallskip\noindent\textbf{EDT recommendation requires new foundations.}
\emph{Why are existing solutions in data systems or recommendation systems
insufficient?} EDT recommendation systems lie at the intersection of
traditional (non-exploratory) data systems and content recommendation
systems~\cite{abdollahpouri2017controlling, baeza2004query, gomez2015netflix,
gupta2014real, grbovic2015commerce}, inheriting challenges from both
(Table~\ref{tab:edt-comparison}). Traditional data systems typically assume
precise specifications and lack support for personalization. Recommendation in
data systems mostly focused on bespoke solutions for individual tasks, such as
visualization~\cite{ DBLP:journals/pvldb/VartakRMPP15} and
exploration~\cite{DBLP:conf/icde/JoglekarGP16}, but they do not generalize to
other EDTs or support multiple recommendations.

\looseness-1 While personalized recommendation systems have seen significant
advances in content domains---e.g., documents in search
engines~\cite{abdollahpouri2017controlling, baeza2004query}, media in streaming
platforms~\cite{gomez2015netflix} and social media~\cite{gupta2014real}, and
products in e-commerce~\cite{grbovic2015commerce}---their adaptation to EDTs
poses two computational challenges. \emph{First}, content recommendation
systems typically assume that the candidate space is pre-enumerated, such as a
catalog of movies. In contrast, EDTs often involve combinatorially large,
programmatically generated spaces of possible \emph{actions} that are not
stored explicitly (Example~\ref{ex:one}). \emph{Second,} they assume that all
candidates (or their embeddings) are fully observable and can be efficiently
scored against a user-preference function. In contrast, assessing the relevance
of EDT actions requires (i)~costly materialization of candidate actions over
the specific data instance (e.g., executing a query over a database) and then
(ii)~measuring the relevance of the materialized results (e.g., output of a
query). In essence, content recommendation systems suggest the data itself,
whereas EDTs need recommending \emph{actions}---programs that are applied to
data.

\begin{table}[t]
\centering
\small
\resizebox{0.99\columnwidth}{!}{
\setlength{\tabcolsep}{2pt}
\renewcommand{\arraystretch}{0.9}
\begin{tabular}{@{}lc@{ }c@{\phantom{x}}c@{}}
\hline
\textbf{Requirements \& properties} &
\rotatebox{360}{\makecell{Traditional\\Data Systems}} &
\rotatebox{360}{\makecell{Content Rec.\\Systems}} &
\rotatebox{360}{\makecell{\textbf{EDT Rec.}\\\textbf{\phantom{s}Systems}}} \\
\midrule
User-oriented: personalized \& adaptive      &            & \checkmark & \checkmark \\
\rowcolor{gray!15}
Subjective goal \& vague success metric	 	 &			  & \checkmark & \checkmark \\
Context matters                              &            & \checkmark & \checkmark \\
\rowcolor{gray!15}
Novelty/serendipity matters                  &            & \checkmark & \checkmark \\
Semantic understanding required		         &            & \checkmark & \checkmark \\
\rowcolor{gray!15}
Interactive speed required		     		 &            & \checkmark & \checkmark \\
Combinatorial candidate space                & \checkmark &            & \checkmark \\
\rowcolor{gray!15}
Dependency among candidates               & \checkmark &            & \checkmark \\
Search over sets of candidates               & \checkmark &            & \checkmark \\
\rowcolor{gray!15}
Utility unknown before materialization     	 & \checkmark &            & \checkmark \\
High materialization cost	 				 & \checkmark &            & \checkmark \\
\bottomrule
\end{tabular}
}
\caption{\small EDT recommendation inherits challenges from traditional data systems
and content recommendation systems.}

\vspace{-9mm}
\label{tab:edt-comparison} 
\end{table}

\vspace{0.5mm}\noindent\textbf{LLMs and AI agents: new enablers for EDT
recommendation.} The recent development of Large Language Models (LLMs) and AI
agents makes our vision timely and the unique challenges in EDT recommendation
tractable. Specifically, LLMs and AI agents enable development of effective
techniques in ways that were not previously possible: (1)~LLMs can provide
semantic understanding of data to estimate action utility from structural and
domain-specific semantics, enabling effective pruning of candidate-action space
and materialization cost reduction; and (2)~AI agents can track inter-task
dependencies and fetch external information, which enables coordinated
recommendation across tasks.

However, these capabilities raise a fundamental question: \emph{Can EDT
recommendation be solved solely using LLMs and AI agents?} Despite their
semantic reasoning and planning capabilities, these systems primarily
\emph{generate} plausible but unvalidated actions~\cite{cho2025data}. However,
EDT recommendation is fundamentally a \emph{search and optimization problem,
not a generation problem}. Even AI agents remain bottlenecked by the cost of
search and validation~\cite{DBLP:conf/icml/KambhampatiVGVS24,
DBLP:conf/iclr/StechlyVK25}. Thus, EDT recommendation requires new foundations,
with LLMs and agents as enablers, not complete solutions.

\rev{\vspace{0.5mm}\noindent\textbf{AI agents as a new class of EDT users.} AI
agents are also becoming primary \emph{users} of data systems. Without
guidance, an agent brute-forces the action space by repeatedly generating,
executing, and inspecting candidate actions. Instead, the agent can call an EDT
recommender with its intent and context, and receive validated actions ranked
by estimated utility, reducing the actions the agent executes and improving
coverage of the action space.}

\vspace{0.5mm}\noindent\emph{Organization.} The rest of the paper is organized
as follows. Section~\ref{sec:variants} presents a categorization of EDT
recommendation systems along two axes---recommendation modality and task
setting. Section~\ref{sec:design} describes the core components of EDT
recommendation systems. Section~\ref{sec:agenda} outlines a research agenda
based on the key challenges to building theoretical foundations and ensuring
the practical requirements of EDT recommendation systems.
Section~\ref{sec:evaluation} presents an evaluation framework.
Section~\ref{sec:related} positions our vision with respect to prior work.
Section~\ref{sec:conclusions} provides concluding remarks.


\section{\hspace{-2mm}EDT recommendation system variants}\label{sec:variants}

In this section, we informally formulate the problem of EDT action
recommendation and categorize EDT recommendation systems along two
axes---recommendation modality and task setting.

\newtheorem{problem}{Problem}

\vspace{-1mm}
\begin{problem}[EDT Action Recommendation]
For a specific EDT, given a user, a data instance, constraints, and context,
the EDT action recommendation problem is to identify the best action(s) that
maximize progress toward the user's (implicit) exploratory goals.
\end{problem}
\vspace{-1mm}

\rev{\noindent We envision the EDT recommender as a middleware layer in data
systems, between front ends and the database engine. It reuses the query
optimizer (statistics, cost estimates, and materialized views) to estimate and
validate actions cheaply, and leverages a semantic layer (e.g., LLM-derived
semantics) to assess action utility.}

\begin{figure}[t]
	\vfill
    \centering
    \includegraphics[width=0.92\columnwidth]{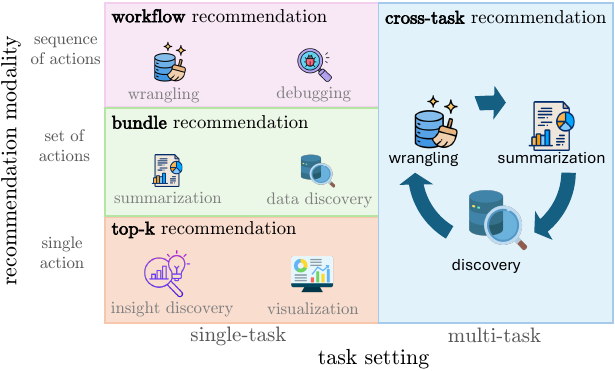}
	\vspace{-4mm}
    \caption{\small EDT recommendation spans two task settings and three recommendation modalities.}
	\vspace{-4mm}
    \label{fig:overview}
\end{figure}

\looseness-1 Figure~\ref{fig:overview} shows our categorization of EDT
recommendation along two axes: task settings---\emph{single-task}, where
recommendations support one EDT, and \emph{multi-task}, where recommendations
span multiple EDTs---and recommendation modalities: \emph{single-action}, where
top-$k$ actions are recommended, such as insight
discovery~\cite{QuickInsightsDing19} and
visualization~\cite{DBLP:journals/pvldb/VartakRMPP15}; \emph{bundle}, where a
set of complementary actions is recommended together, such as data
summarization~\cite{cho2025data} and
discovery~\cite{DBLP:journals/pvldb/RezigBFPVGS21}; and \emph{workflow}, where
a sequence of actions is recommended, such as data
wrangling~\cite{DBLP:conf/sigmod/ChopraFGHPRSS023} and
debugging~\cite{DBLP:conf/cidr/RezigCSSMTOS20}. This categorization yields four
variants of EDT recommendation systems.

\smallskip\noindent\textbf{Top-$k$ recommendation for single task.}
\looseness-1 The simplest setting for EDT recommendation is single-action
recommendation within a single-task setting, where users execute one action at
a time and the outcome of each action informs next exploration steps. Examples
include insight discovery, visualization, data profiling, dataset search,
hypothesis exploration, and anomaly discovery. For these tasks, the
recommendation system should rank candidate actions by their utility to the
user and return the \emph{top-$k$} actions, allowing users to choose among
multiple promising next actions. Notably, solutions for this setting provide
the foundation for more complex settings that require recommending multiple
actions or spanning multiple EDTs.

Despite their diverse objectives, recommendations across EDTs share a set of
fundamental requirements (Table~\ref{tab:edt-comparison}, \S\ref{sec:agenda}).
Thus, instead of task-specific bespoke recommendation systems, as exemplified
by existing systems in Table~\ref{tab:edtcomp}, a \emph{generalizable
framework} should be developed for top-$k$ EDT action recommendation that can
be instantiated for any single EDT. The framework should factor out shared
components across EDTs, such as search-space reduction and fast
candidate-action materialization, to ensure their reuse. It should require only
two task-specific components: (i)~a language of candidate actions, defined by
its vocabulary, grammar, and operational semantics on the data (how to execute
an action over a dataset), and (ii)~a utility notion to score the materialized
results.

\smallskip\noindent\textbf{Bundle recommendation for single task.} Many EDTs
involve a \emph{set} of actions~\cite{zhu2014bundle} that must jointly optimize
a global objective---such as maximizing data coverage during data
exploration~\cite{DBLP:conf/icde/JoglekarGP16}---or satisfy global
constraints---such as ensuring $\text{diversity}$ among data
summaries~\cite{cho2025data}. While general constrained set discovery with a
global objective is NP-hard, prior solutions for bundle recommendation in
data-centric tasks~\cite{ranu2014answering, DBLP:conf/icde/JoglekarGP16,
10.14778/3192965.3192969, DBLP:journals/pacmmod/FanHXZ24,
DBLP:journals/pvldb/KrishnanWWFG16, 10.1145/3654969, blau2025causal,
jin2020auto, cho2025data, DBLP:conf/icdt/Moumoulidou0M21} exploit properties of
the constraints or objectives, such as submodularity, to devise practically
efficient approximate solutions. However, these solutions are often
task-specific and do not generalize across EDTs.

\looseness-1 Unlike bundle recommendation over standalone
items~\cite{kunaver2017diversity}, EDT actions require materialization over
specific data instances. For example, consider the global constraint of
ensuring diversity by requiring the distance between every pair of recommended
items to exceed a threshold. The distance between two standalone items $c_1$
and $c_2$ can simply be measured as $dist(E(c_1), E(c_2))$, where $E$ maps each
item to an embedding space. In contrast, given a database $D$, the distance
between two actions $a_1$ and $a_2$ must be computed over their materialized
results, $dist(E(a_1(D)), E(a_2(D)))$, where each action $a_i$ is parameterized
by $D$. As such, bundle recommendation for EDTs must strategically \emph{push
down} global objectives and constraints into the recommendation process, in the
same spirit as predicate pushdown~\cite{selinger1979access,
hellerstein1993predicate, ullman1988principles, levy1994query,
yan2023predicate} in query optimization.

\smallskip\noindent\textbf{Workflow recommendation for single task.}
\looseness-1 For EDTs that require a sequence of interdependent actions,
decisions made early in the workflow constrain later choices, which makes
workflows fundamentally different from unordered bundles. Finding the highest
utility workflow for recommendation is a sequence optimization problem, similar
to planning, scheduling, and job sequencing. However, since each EDT action
transforms the data, subsequent actions operate on the resulting intermediate
data. For example, during wrangling, recommending (i)~conversion of risk
probability to a discrete ordinal scale of 1--5 (for categorical analysis),
followed by recommending (ii)~missing value imputation using the attribute
mean, introduces incompatible values w.r.t.\ the ordinal scale (e.g., 3.36). This
calls for targeted techniques, such as leveraging semantics of actions, to
ensure compatibility within the recommended workflow.

\smallskip\noindent\textbf{Unified recommendations across tasks.} \looseness-1
A key characteristic of EDTs is their interconnected and interleaved nature in
a multi-task setting. Exploratory data analysis may begin with discovery to
identify relevant data, proceed through wrangling to clean and reshape the
data, and conclude with summarization to extract insights. A one-shot workflow
fails in this setting because the process is iterative and actions are mutually
dependent. For instance, insights from summarization often trigger further
discovery, and newly discovered data may require additional wrangling.
Recommending for each EDT in isolation fails to account for how recommendations
in one task influence recommendations in another, calling for coordination
across recommenders (Example~\ref{ex:three}).


\section{EDT recommendation Components}\label{sec:design}

EDT recommendation can be approached through three paradigms. The first is
\emph{generation-based}, where generative AI directly constructs a plausible
action without considering all possible alternatives. However, such
recommendations are unvalidated, providing no guarantee that the recommended
action has high utility w.r.t.\ the data instance. The second is
\emph{prediction and ranking based}, prevalent in traditional recommendation
systems, where an ML model assigns utility scores to candidate EDT actions
based on materialized results. This approach requires candidate actions to be
materialized beforehand over the data, and it also requires training data,
which is often unavailable in practice. The third paradigm models EDT
recommendation as a \emph{search and optimization} problem that leverages the
structure of the problem and properties of the utility model and constraints:
only promising candidate regions are explored, the utility of materialized
results can be approximated, and efficient approximate search algorithms can be
devised. We argue that this search-and-optimization paradigm is well suited to
EDT recommendation because it directly addresses the combinatorial action space
and costly materialization issues inherent to EDTs.

\begin{figure}[t]
    \centering
    \includegraphics[width=0.82\columnwidth]{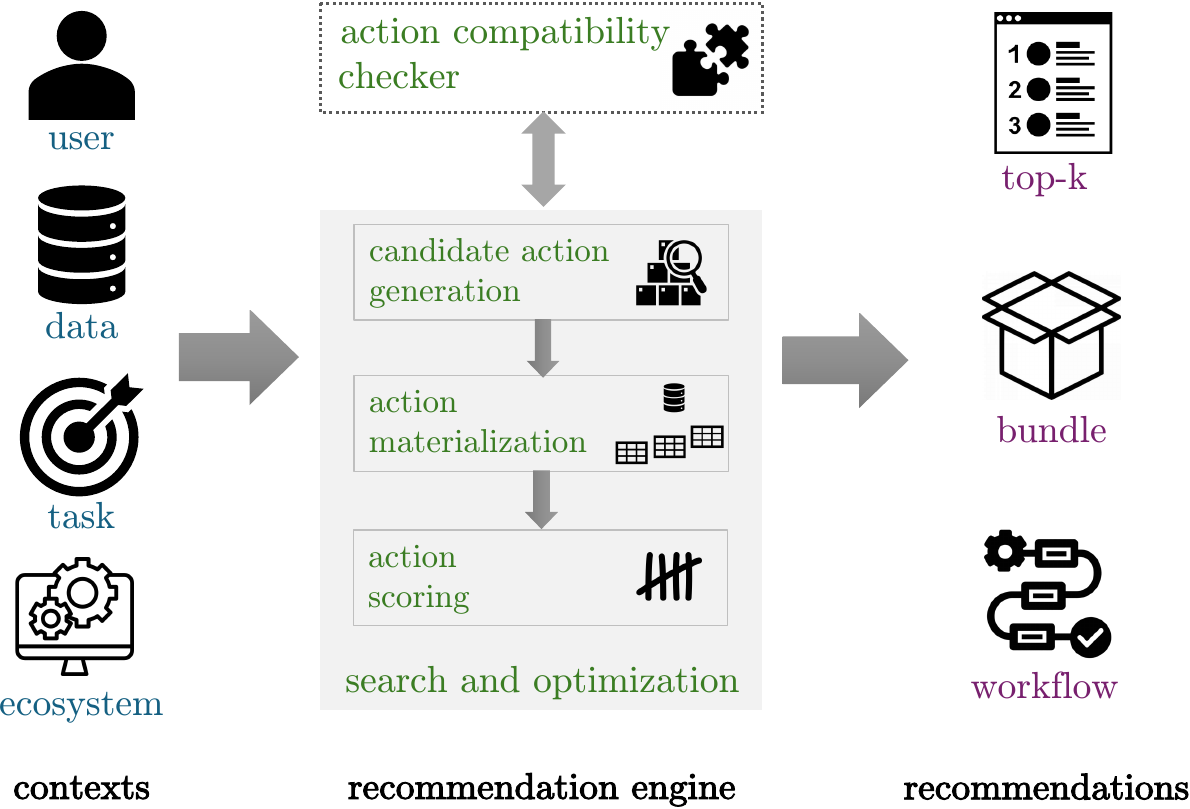}
	\vspace{-3mm}
	\caption{\small Core components of an EDT recommendation system.}
	\vspace{-2mm}
	\label{fig:corecompo}
\end{figure}

\smallskip\noindent\textbf{Overview.} Figure~\ref{fig:corecompo} shows the core
components of a generalizable EDT recommendation system within this paradigm.
Central to EDT recommendation across all variants are personalization and
context-awareness, requiring the system to model user, data, task, and
ecosystem \emph{contexts} that serve as input to the \emph{recommendation
engine}. The recommendation engine consists of a \emph{search and optimization}
module, which efficiently explores the combinatorial action space while
reducing the materialization cost required to score candidates, and an
\emph{action compatibility checker}, which enforces global constraints and
ensures compatibility among recommended actions. The output of the
recommendation engine is either \emph{top-$k$} actions, a \emph{bundle} of
actions, or a \emph{workflow} (sequence of actions).

\subsection{Contexts} \looseness-1 Four sources of context are relevant for EDT
action recommendation: (i)~\textbf{user}: intent, preferences, historical
behavior, and feedback; (ii)~\textbf{task}: the primary goal (e.g.,
summarization), secondary goals (e.g., interpretability), and constraints
(e.g., diversity); (iii)~\textbf{data}: semantics, metadata, and profiles; and
(iv)~\textbf{ecosystem}: workflow, tools in use, and search and communication
history. Table~\ref{tab:edtcomp} shows examples of established and emerging
EDTs along with representative works, which provide limited support for these
contexts.

\begin{example}[Leveraging context for unified recommendations]
	 A data scientist begins by cleaning sensor data, where the recommendation
	 system suggests forward-filling imputation for missing temperature readings,
	 leveraging its time-series nature (data context). During feature selection,
	 the system avoids recommending actions that rely on temperature, recognizing
	 its reduced reliability due to imputation (workflow context). During
	 visualization, it suggests binning temperature into wide intervals to reduce
	 noise from imputation (task context) and recommends bar plots, drawing on
	 user preferences observed in other documents (user context).
\label{leverage}
\end{example}

\vspace{-1mm}

As Example~\ref{leverage} highlights, user context, such as response to earlier
recommendations, enables \emph{personalization} and \emph{adaptivity}. Task
context provides guidance to model action utility and properties that can guide
search and optimization, such as prioritizing task-relevant actions. Data or
domain context, which can be obtained from LLMs, provides essential signals for
semantic understanding and action utility estimation. Ecosystem
context---including recommendations made in other tasks, how prior actions
transformed the data, the software environment, search history, and chat or
email interactions---enables cross-task coordination.

\looseness-1 
\begin{table}[t]
\newcommand{\crossmark}{$\mathbf{\times}$}
\centering
\resizebox{\columnwidth}{!}{
\begin{tabular}{@{}c@{\phantom{x}}llc@{}c@{}c@{\phantom{x}}c@{}}
\toprule
& \multirow{2}{*}{\textbf{Task}} & \multirow{2}{*}{\textbf{\makecell{Representative\\work}}}  & \multicolumn{4}{c}{\textbf{Contexts}} \\ 
\cline{4-7} & & & \makecell{\textbf{user}\\[-4pt]\phantom{x}} & \makecell{\textbf{data}\\[-4pt] \textbf{semantics}} & \makecell{\textbf{task}\\[-4pt]\phantom{x}} & \makecell{\textbf{ecosystem}\\[-4pt]\phantom{x}} \\
\midrule
\multirow{5}{*}{\rotatebox{90}{\small \textbf{Established}}}
& Data exploration & Smart Drill-Down~\cite{DBLP:conf/icde/JoglekarGP16}  & \crossmark & \crossmark & \crossmark & \crossmark\\
& Data preparation & Auto-Suggest~\cite{DBLP:conf/sigmod/YanH20}  &  \crossmark &  \crossmark & \checkmark & \crossmark\\
& Data summarization & SAGE~\cite{cho2025data} &  \checkmark &  \checkmark & \crossmark & \crossmark\\
& \multirow{2}{*}{Data visualization} & SeeDB~\cite{DBLP:journals/pvldb/VartakRMPP15}  & \crossmark & \crossmark & \crossmark & \crossmark\\
& & ShiftScope~\cite{DBLP:conf/sigmod/SahaABT24}  & \checkmark & \crossmark & \checkmark & \crossmark\\
\midrule
\multirow{4}{*}{\rotatebox{90}{\small \textbf{Emerging}}} 
& Cherry-picked generalizations & OREO~\cite{DBLP:journals/pvldb/LinYMJM22}  & \crossmark & \crossmark & \crossmark & \crossmark\\
& Explain data change & Explain Da-V~\cite{DBLP:journals/pvldb/ShragaM23}  & \crossmark & \crossmark & \crossmark & \crossmark\\
& Explain outliers in aggregates & CAPE~\cite{DBLP:conf/sigmod/MiaoZGR19}  & \crossmark & \crossmark & \crossmark & \crossmark\\
& Discover entity-enhancing rules & Fan et al.~\cite{DBLP:journals/pacmmod/FanHXZ24}  & \checkmark & \crossmark & \crossmark & \crossmark\\
\bottomrule
\end{tabular}}
\caption{\small Examples of EDTs and representative work.}\label{tab:edtcomp}
\vspace{-7mm}
\end{table}

\vspace{-1mm}

\subsection{Recommendation engine}

\subsubsection{Search \& optimization} \looseness-1
This module is responsible for efficiently searching for high-utility EDT
actions, which requires managing the large and expensive-to-materialize
candidate action space. It generates promising candidates by prioritizing those
that align with the user's preferences and task requirements. It further
reduces unnecessary materialization by pruning irrelevant, semantically
invalid, or low-potential candidates using available context.

For bundle and workflow recommendation, this module searches over combinations
and sequences of actions while enforcing global constraints, such as diversity
and consistency. It pushes properties of these constraints into candidate
generation and materialization, pruning unpromising candidates that cannot
contribute to a high-quality bundle or valid workflow. Full materialization is
only pursued for the most promising candidates, striking a balance between
maintaining high recommendation quality and ensuring low recommendation latency.

\subsubsection{Action compatibility checker} \looseness-1
This module ensures that recommended actions are coherent and compatible,
especially for workflows, where each action transforms the data and affects
subsequent actions that operate on the transformed data. It tracks order
dependencies among actions and eliminates candidates that yield incompatible
sequences. Further, unified recommendation requires coordination among
task-specific recommenders, tracking inter-task dependencies, and obtaining
relevant external information.


\section{Research Agenda}\label{sec:agenda}

EDT recommendation systems combine requirements from traditional data systems
and recommendation systems (Table~\ref{tab:edt-comparison}), requiring
personalization, adaptation, semantic understanding, and interactive response
times. However, the combinatorial action space and the unknown utility of
actions until materialization on data pose efficiency challenges. Below, we
outline ten research directions toward realizing our vision of EDT
recommendation systems.

\smallskip\noindent\textbf{Learn user preferences over action space.}
Personalization requires modeling user preferences and intent, which are often
implicit and must be learned from interactions~\cite{koren2008personalized,
resnick1997recommender}. The combinatorial action space makes learning a
complete user--action relevance matrix infeasible. However, EDT actions are
compositions of a much smaller set of \emph{primitives}, including operations
(e.g., \texttt{FILTER} or \texttt{SPLIT} for wrangling) and their parameters,
such as data attributes, aggregate functions, and constants. Thus, a promising
direction is to learn user relevance models over these primitives and compose
these learned preferences to estimate action relevance.

\smallskip\noindent\textbf{Model rich data context.} Prior
work~\cite{DBLP:journals/pvldb/VartakRMPP15, HarrisWWW23SpotLight,
DBLP:journals/pvldb/DemiralpHPP17, QuickInsightsDing19} primarily used
statistics and metadata to model data context. However, data semantics provide
richer signals, which can flag invalid actions (e.g., sum of ZIP codes) and
distinguish expected from surprising insights (Example~\ref{ex:semantics}).
LLMs can be utilized to infer such \emph{semantic context} by estimating the
probabilities of domain-specific hypotheses.

\vspace{-1mm}
\begin{example}
    Outliers can signal valuable insights---e.g., \texttt{PhD} holders in
    \texttt{IT} earning $\$140$K vs.\ a general average of $\$60$K. However,
    not all outliers are insightful; some, like this, are expected. In
    contrast, high blood pressure in kids indicates a genuine insight.
\label{ex:semantics}
\end{example}
\vspace{-1mm}


\smallskip\noindent\textbf{Search space reduction and search acceleration.}
\looseness-1 Possible directions to address the combinatorial search space
include: (1)~restrict candidate generation to operations and parameters
relevant to the user, data, and task contexts, e.g., prioritize user-preferred
operations and omit cryptic attributes like \texttt{job\_code} when
interpretability is desired; (2)~prune unpromising candidates based on action
semantics, such as avoiding ``drop rows with any missing value'' when 90\% of
values are missing; (3)~bias candidate generation by pushing down global
constraints, e.g., avoid similar actions when diversity is desired; and
(4)~cluster actions across multiple dimensions, such as user preferences,
action semantics, and data impact, to focus search on relevant clusters and
promote bundle diversity by discouraging multiple selections from the same
cluster.

\smallskip\noindent\textbf{Action materialization cost reduction.} \looseness-1
Beyond established tech\-niques---such as approximate query processing
\cite{park2018verdictdb}, sampling~\cite{babcock2003dynamic}, distributed
computing~\cite{dean2008mapreduce}, and lazy materialization
\cite{zhou2007lazy}---LLMs could estimate action utility from action semantics,
data statistics, and context. For example, an LLM could estimate filter result
size from its condition and data statistics, avoiding materialization. Other
directions include anytime materialization~\cite{raman2002partial} that
terminates early for unpromising actions and reinforcement learning to predict
data-specific utility without materialization. For workflows, materialization
could also terminate as soon as incompatibility w.r.t.\ actions selected earlier
is detected. E.g., if earlier actions rely on a functional dependency, a
candidate action can be discarded once it modifies the data in a way that
violates that dependency.

\smallskip\noindent\textbf{Decompose global objectives for algorithmic
benefits.} Recommending a set of actions that maximize a global objective is
generally NP-hard~\cite{sakai2003note}. Certain decomposability properties of
global objectives, such as submodularity and monotonicity, admit efficient
greedy solutions with approximation guarantees. Thus, a promising direction is
to develop a taxonomy of decomposable global objectives across EDTs,
characterize their algorithmic properties, and identify approximation
algorithms that generalize across EDTs.

\smallskip\noindent\textbf{Decompose global constraints for computational
benefits.}
Learning ``action embeddings'' could enable decomposition of concepts such as
distance, used in global constraints like diversity, to accelerate computation
(Example~\ref{ex:24}). Action embeddings should capture three aspects:
(1)~\emph{structural embedding}, representing action-defining elements such as
selection predicates or operations like \texttt{fillna}; (2)~\emph{semantic
embedding}, capturing what the action does, such as compression, addition, or
aggregation; and (3)~\emph{impact embedding}, representing the action's effect
on the data.

\begin{example}[Decomposing distance] \label{ex:24} \looseness-1 
    ``Distance'' can be decomposed into 3 components: (D1-cheapest) structural
    disjointness of actions, (D2-moderate) data-agnostic semantic difference of
    actions, and (D3-costliest) divergence in outcomes when actions are applied
    to data. Under diversity constraints, an action can be skipped if it fails
    the D1 check w.r.t.\ the already selected actions.
\end{example}

\noindent\textbf{Model action compatibility.}
\looseness-1 Action compatibility can be modeled as a weighted directed graph,
but constructing it over the combinatorial action space is infeasible. A
promising direction is building a task-specific \emph{compatibility graph} over
action clusters. If no edge connects clusters $A$ and $B$, no action in $A$ can
precede one in $B$ (e.g., lossy transformations cannot precede imputation).

\smallskip\noindent\textbf{Adaptive recommendation.} Adaptive recommendation
\cite{machado2021aware} can incorporate user feedback by modeling the semantics
of user interactions with recommended actions. For instance, modifying a
filter's threshold signals relevance to the operation but not its parameters. A
key challenge is balancing recent feedback with historical behavior. A
direction is to learn \emph{temporal user preference embeddings} from edits,
refinements, and rejections, capturing short- and long-term preferences through
learnable \emph{temporal decay}.

\smallskip\noindent\textbf{Multi-agent collaboration for cross-task
coordination.} Agentic approaches enable agents to acquire relevant external
information and coordinate across tasks~\cite{zhu2025multiagentbench,
grotschla2025agentsnet}. Multi-agent
collaboration~\cite{talebirad2023multi,dorri2018multi} could serve as an
\emph{outer orchestration layer}, with per-task recommenders providing
guardrails through (i)~a \emph{context-request} interface for missing context
and (ii)~a \emph{conflict-report} interface to help resolve cross-task
incompatibilities.

\rev{\smallskip\noindent\textbf{Agent-facing recommendation.} Serving agents as
users raises new challenges. First, the tool should expose machine-readable
actions with utility estimates, and accept agent-specified budgets and
constraints. Second, agents issue frequent bursts of requests, which calls for
sharing computation across requests. Third, agent feedback is abundant but
noisy, which requires additional calibration.}


\section{Evaluation Framework}\label{sec:evaluation}

Evaluating EDT recommendation systems is challenging due to their user-centered
nature and the lack of benchmark datasets in this space, which necessitates a
multi-pronged evaluation strategy.

\smallskip\noindent\textbf{Evaluation axes.} EDT recommendation systems should
be evaluated along six axes using established and new metrics.

\vspace{0.5mm}\noindent\emph{\underline{Generalizability.}} Generalizability
across tasks and domains should be assessed by evaluating across diverse EDTs
and data domains.

\vspace{0.5mm}\noindent\emph{\underline{Scalability.}} Scalability should be
evaluated under varying data complexity in \#tuples \& \#attributes and
candidate action space sizes.

\vspace{0.5mm}\noindent\emph{\underline{Efficiency.}} Practical efficiency and
fitness for interactive use should be assessed via recommendation latency and
reduction of search space and materialization cost.

\vspace{0.5mm}\noindent\emph{\underline{Validity.}} Syntactic validity should
be measured by error-free execution of the recommended actions. Semantic
validity should be measured using (i)~precision@k, recall@k, and Mean
Reciprocal Rank (MRR) for top-$k$, (ii)~utility and degree of constraint
satisfaction for bundles, and (iii)~utility and inter-action coherence for
workflows. For the multi-task setting, evaluation should also consider success
rate in achieving the end goal.

\vspace{0.5mm}\noindent\emph{\underline{User alignment.}} Since EDT
recommendation targets human users, it must be validated via user studies,
including controlled experiments, interviews, and contextual observation using
think-aloud protocols. Metrics should include increase in user satisfaction
over EDT iterations to measure \emph{adaptivity} and across sessions to measure
\emph{personalization}. Likert-scale questions such as ``How frequently did you
find the recommended actions relevant?'' can quantify user satisfaction, while
\#iterations and time to task completion can measure the impact of
recommendations on user efficiency. \rev{For agents, the impact should be
measured by comparing agents that leverage the recommender against agents that
do not, in terms of execution cost, token usage, and task success rate.}

\vspace{0.5mm}\noindent\emph{\underline{Robustness and parameter sensitivity.}}
Robustness should be evaluated under noise, data skew, intent ambiguity, and
task complexity. Sensitivity to system parameters, optimization techniques, and
materialization strategies should be analyzed via ablation studies.

\smallskip\noindent\textbf{Human-validated utility models for evaluation at
scale.} \looseness-1 Validating recommendations by obtaining accept/reject
feedback via user studies is prohibitively expensive and cognitively demanding:
without assessing the utility of the alternatives, users cannot correctly judge
recommendation fitness. A solution is to use \emph{human-validated utility
models} that capture real-user preferences and objectively quantify
recommendation quality~\cite{cho2025data}. This involves (i)~devising utility
models covering various EDT-specific criteria such as relevance, novelty,
diversity, coverage, etc.; (ii)~a small-scale user study over several use cases
to identify the best model; and (iii)~using the validated model to evaluate
recommendations at scale.

\vspace{0.5mm}\noindent\textbf{Benchmark datasets.} \looseness-1 Benchmark
datasets should be created in three ways: (i)~workflow extraction from existing
sources such as notebooks~\cite{DBLP:journals/corr/abs-2409-10635};
(ii)~collection of real workflows from user studies over prototypes; and
(iii)~brute-force search to identify ground truths: actions that should be
recommended based on user-validated utility models. For personalization and
adaptivity evaluation, a direction is to create a benchmark with a list of
intents (Example~\ref{ex:one}), and obtain ground-truth workflows from domain
experts.


\section{Related Work} \label{sec:related}

Existing suggestion systems target specific tasks such as data
visualization~\cite{DBLP:conf/sigmod/SahaABT24, DBLP:conf/sigmod/BarbosaMLO12,
DBLP:conf/kdd/QianRDKKMLC21, DBLP:conf/icde/JiLB23,
DBLP:journals/pvldb/DemiralpHPP17, DBLP:journals/pvldb/LeeTABCKMSYHP21,
DBLP:journals/pvldb/VartakRMPP15, DBLP:conf/icde/EhsanSC16,
DBLP:conf/sigmod/KeyHPA12}, exploration~\cite{10.14778/3476311.3476352,
8509239, 10.1007/s00778-013-0311-4, 10.1145/2063576.2063798,
10.1145/3318464.3389779, 10.1145/3448016.3452762, 10.1007/978-3-642-02279-1_2,
DBLP:conf/vldb/Sarawagi00, DBLP:conf/edbt/LipmanMSWZ25},
wrangling~\cite{DBLP:conf/sigmod/ChopraFGHPRSS023, 9953543,
DBLP:journals/corr/abs-2603-21310}, analysis~\cite{DBLP:conf/sigmod/LeeQKO21,
REACT_ContextSensitive_2016}, explanation~\cite{blau2025causal,
blau2025causaldemo}, summarization~\cite{DBLP:journals/pvldb/ChoF24,
cho2025data}, etc. Some focus on suggesting the mechanism (e.g.,
queries)~\cite{10.1007/978-3-642-03730-6_36, 10.1145/3077136.3080652,
DBLP:journals/pvldb/FarihaM19, 5767843, DBLP:conf/edbt/LaiZMACP23, 9767552,
10.14778/1880172.1880175, 10.1145/3318464.3389779}, while others focus on the
results (e.g., tuples). Some consider user profiles, while others leverage
prior query logs. RecDB~\cite{DBLP:journals/pvldb/SarwatAM13,
DBLP:conf/icde/SarwatMMA17} supports personalized data suggestion but does not
suggest the underlying query. ATENA~\cite{10.1145/3318464.3389779} generates an
entire EDA session but without keeping a human in the loop. Suggestion engines
have also been explored for other data-management tasks such as database tuning
and indexing~\cite{7495648, 10.14778/3352063.3352129, 10.14778/2140436.2140444,
839397}, clustering~\cite{DBLP:conf/sigmod/BudalakotiZWKKW24}, time-series
generation~\cite{DBLP:journals/pvldb/AngBHTH24}, identifying valuable
attributes in OLAP~\cite{DBLP:journals/pacmmod/ChenZHW23}, view
generation~\cite{10.14778/1687553.1687556}, and data-science pipeline
construction~\cite{10.14778/3554821.3554882}. However, none of these works
offers a framework that generalizes across EDTs.

\looseness-1 Some approaches use machine learning, such as reinforcement
learning~\cite{DBLP:conf/sigmod/SahaABT24} or convolutional neural
networks~\cite{DBLP:conf/icde/JiLB23}, which assume small candidate space. Most
require materializing all candidates to evaluate their
relevance~\cite{DBLP:conf/kdd/QianRDKKMLC21} and support only top-$k$
suggestions~\cite{DBLP:conf/kdd/QianRDKKMLC21, DBLP:conf/icde/JiLB23,
DBLP:journals/pvldb/DemiralpHPP17, DBLP:journals/pvldb/VartakRMPP15,
DBLP:conf/icde/EhsanSC16}. For suggesting spreadsheet formulas and
data-preparation steps, Auto-Formula~\cite{DBLP:journals/pacmmod/ChenHCFGZ0C24}
and Auto-Suggest~\cite{DBLP:conf/sigmod/YanH20} use machine learning to predict
top-$k$ suggestions. Generally, machine-learning models treat suggesting the
next EDT action as a standard prediction problem. A vast literature exists for
content recommendation~\cite{RUCAIBox_AwesomeRSPapers_2025}, spanning multiple
communities, from information retrieval and recommender systems to data mining,
artificial intelligence, and machine learning. However, adapting content
recommendation systems to EDTs is nontrivial.

In summary, prior works have several limitations: (1)~they target a specific
task and offer a bespoke solution; (2)~they often ignore signals such as the
user's workflow and ecosystem context, failing to align with the user intent;
and (3)~they are history-driven, relying heavily on ML-based approaches akin to
traditional recommendation engines, which are ineffective for EDTs.

\section{Conclusions}\label{sec:conclusions}

\looseness-1 We presented a vision of recommendation systems for exploratory
data tasks where recommendation capability is a first-class citizen of data
systems, not an afterthought or a superficial chatbot feature. We characterized
EDT recommendation across single-task and multi-task settings and
single-action, bundle, and workflow modalities. We provided a design of a
generalizable and unified recommendation framework, outlined a research agenda
spanning context modeling, search-space and materialization cost reduction, and
cross-task coordination, and devised an evaluation framework. Realizing this
vision requires addressing the unique challenges of EDT recommendation by
integrating techniques from data management, recommendation systems, and
constrained optimization, with capabilities of large language models and AI
agents. Our vision is to make action recommendations for EDTs more trustworthy,
effective, and accessible, democratizing both existing and emerging tasks,
\rev{for human users and AI agents,} which can ultimately empower millions of
analysts, scientists, engineers, journalists, and policymakers to perform
complex exploratory data tasks more accurately and efficiently in diverse
real-world settings.


\clearpage

\bibliographystyle{ACM-Reference-Format}
\bibliography{paper}

\end{document}